\newif\ifsubmission
\submissiontrue
\ifsubmission
\documentclass[sigconf]{acmart}
\else 
\documentclass[sigconf, anonymous, review]{acmart}
\fi

\newif\ifpreprint
\preprinttrue

\usepackage[nolist,nohyperlinks]{acronym}
\usepackage{listings}
\usepackage{algpseudocode}
\usepackage{algorithm}
\usepackage{tabularx}
\usepackage{natbib}
\usepackage{listings}
\usepackage{layouts}
\usepackage{xcolor}
\usepackage{siunitx}
\usepackage{caption}
\usepackage{subcaption}
\usepackage[inline]{enumitem}
\usepackage[noabbrev, capitalise]{cleveref}

\AtBeginDocument{%
  \providecommand\BibTeX{{%
    \normalfont B\kern-0.5em{\scshape i\kern-0.25em b}\kern-0.8em\TeX}}}

\ifpreprint
\renewcommand\footnotetextcopyrightpermission[1]{} 
\else 
\setcopyright{acmcopyright}
\copyrightyear{2022}
\acmYear{2022}
\acmDOI{XXXXXXX.XXXXXXX}

\acmConference[WiSec'22]{}{May 16--19,
  2022}{San Antonio, Texas, USA}
\fi
\begin{document}

\lstset{
 breaklines=true
 }

\newcommand*{\AG}{\textit{AirGuard}}

\newif\ifanonymous
\anonymousfalse
\newif\ifextended
\extendedfalse
\newif\ifeprint
\eprinttrue

\ifsubmission
	\newcommand{\TODO}[1]{}
\else
	\newcommand{\TODO}[1]{{\color{red}TODO: #1}}
\fi

\title[\AG{}]{AirGuard - Protecting Android Users From Stalking Attacks By Apple Find My Devices}

\author{Alexander Heinrich}
  \affiliation[obeypunctuation=true]{
    \department[0]{Secure Mobile Networking Lab}\\
    \department[1]{Department of Computer Science}\\
    \institution{TU Darmstadt}, \country{Germany}
  }
  \email{aheinrich@seemoo.de}
  \orcid{0000-0002-1150-1922}

\author{Niklas Bittner}
  \affiliation[obeypunctuation=true]{
    \department[0]{Secure Mobile Networking Lab}\\
    \department[1]{Department of Computer Science}\\
    \institution{TU Darmstadt}, \country{Germany}
  }
  \email{nbittner@seemoo.de}
  
\author{Matthias Hollick}
  \affiliation[obeypunctuation=true]{
    \department[0]{Secure Mobile Networking Lab}\\
    \department[1]{Department of Computer Science}\\
    \institution{TU Darmstadt}, \country{Germany}
  }
  \email{mhollick@seemoo.de}
  \orcid{0000-0002-9163-5989}


\begin{abstract}
    Finder networks in general, and Apple's Find My network in particular, can pose a grave threat to users' privacy and even health if these networks are abused for stalking.
    Apple's release of the AirTag---a very affordable tracker covered by the nearly ubiquitous Find My network---amplified this issue. 
    While Apple provides a stalking detection feature within its ecosystem, billions of Android users are still left in the dark. 
    Apple recently released the Android app ``Tracker Detect,'' which does not deliver a convincing feature set for stalking protection. 
    We reverse engineer Apple's tracking protection in iOS and discuss its features regarding stalking detection. We design ``\AG{}''
    and release it as an Android app to protect against abuse by Apple tracking devices. 
    We compare the performance of our solution with the Apple-provided one in iOS and study the use of \AG{} in the wild over multiple weeks using data contributed by tens of thousands of active users. 
\end{abstract}




\maketitle
\ifpreprint
\pagestyle{plain}
\fi 
\TODO{Todos are marked in red}


\section{Introduction}

Apple's release of the AirTag in April 2021 drastically changed the application domain of so-called ``key finders'' or ``trackers,'' i.e., battery-operated devices that can be attached to personal items of importance and allow to locate these items if lost, misplaced, or stolen.

While the first generation of finders by different manufacturers was typically directly linked to the user's smartphone and, hence, limited in range and capabilities, the current finders operate using large-scale finder networks such as Apple's Find My network~\cite{xylabsXYSecureIOS2014,appleinc.AirTag2021}. 

In the case of loss or theft, the owner can receive a detailed location report of the device. 
AirTags are linked to Apple IDs and the owner can then view the location in the Apple Find My app on an iPhone, Mac, or Apple Watch (i.e., linked to the same Apple ID). The accuracy of these reports in practice is around \SI{30}{\metre} in urban environments~\cite{heinrichWhoCanFind2021a}.  

Apple's privacy-first approach to the Find My network delivers leading security and privacy properties for finder networks. It was found to protect the location privacy of the legitimate AirTag owner against outsiders as well as against Apple's ecosystem \cite{heinrichWhoCanFind2021a}.
The security and privacy of many other finder systems was in a dismal state prior to the release of the AirTag \cite{wellerLostFoundStopping2020a}.

However, AirTags can also be used for nefarious purposes. Given the diminutive size of only \SI{3}{\centi\metre}, they can be easily hidden, thus allowing for tracking or stalking unsuspecting victims. The ubiquitous nature of the Find My network, combined with its high accuracy and low entry cost, lowers the bar for abuse. 
The media reacted promptly with articles linking AirTags with domestic abuse and stalking attempts~\cite{macAreAppleAirTags2021, cahnAppleAirTagsAre2021,fowlerReviewAppleAirTag2021}.
Moreover, exposed individuals such as celebrities, activists, or critical journalists might further be targeted by paparazzi, secret services, or oppressive regimes, respectively. 
AirTag abuse is not limited to stalking or domestic abuse, but they are also used to identify the parking location of valuable cars, allowing to steal the cars thereafter ~\cite{macAreAppleAirTags2021,koskiDiscoveryAirTagTracking2021}. 
Some of the known abuse cases were uncovered due to the automatic tracking protection feature offered by Apple devices such as iPhones. The tracking protection identifies suspicious AirTags in the user's surroundings and notifies the user when a device has been following them for a prolonged period~\cite{macAreAppleAirTags2021}.

AirTags can potentially endanger all non-Apple users in the aforementioned abuse scenarios because other smartphone ecosystems like Google's Android do not include compatible and mandatory tracking protection. It took Apple more than six months before releasing an Android app called ``Tracker Detect'' on December 11, 2021, to remedy this issue partially. Unfortunately, this app is ill-suited for the purpose and not usable in practice, as it requires the user to perform repeated manual scans to find a tracking device hopefully (see \cref{sec:background_stalkingProtection}). As a result, Apple's current efforts leave all non-Apple users wide open for abuse. 


Our goal is to provide non-Apple users with comprehensive anti-tracking protection. 
We emphasize automatic operation and prioritize user interface design to support non-experts. 
Lastly, our solution eschews including features that could be used for abuse.  

Our key contributions are: 
\begin{enumerate}
    \item We reverse-engineer Apple's tracking protection in iOS.
    \item We design, build and release the open-source\footnote{\url{https://github.com/seemoo-lab/AirGuard}} Android app \AG{} to protect people from AirTag abuse.
    \item We evaluate \AG{} and compare it against the tracking protection implemented in iOS.
    \item We analyze an anonymous dataset generated by \AG{} users to analyze tracking attacks in the wild.
\end{enumerate}

This paper is structured as follows: \cref{sec:background} introduces the Find My network that powers AirTags and other Find My devices. In \cref{sec:tracking_detection}, we reverse-engineer the tracking detection in iOS. We explain how trackers are detected and when the system notifies the user. In \cref{sec:airguard}, we introduce our app \AG{}. We describe its features, highlighting the user interface design and the tracking detection algorithm. \cref{sec:evaluation} evaluates \AG{} in three scenarios and compares it to the iOS tracking detection algorithm. In \cref{sec:user_study}, we analyze the data from our user study. We present how many Find My devices are in use and how many users of our app have been notified about a potential tracker. 
We conclude our work in \cref{sec:conclusion}.

\section{Background and Related Work} \label{sec:background}

This section describes the essential operation of Bluetooth item finders with a finder network, discusses details of Apple's Find My network, characterizes several stalking protections in place, and introduces work on analyzing Apple wireless services and other Bluetooth item finders.

\subsection{Bluetooth Device Finders}


Many Bluetooth-based item finders, or key finders, are now using finder networks aiding to find lost or misplaced trackers~\cite{tileinc.HowDoesTile2020, appleinc.AppleFindMy2021}. Those networks are generally based on apps from item finder manufacturers. 
Any user of the app helps to find lost or stolen item finders. In principle, whenever an item finder is discovered via Bluetooth with the app, the app automatically reports the current location to the manufacturer, who then sends a notification of the discovered device to the item finder's owner. 
The ability to locate lost devices and the accuracy of reported locations mostly depends on the number of Android app users. iOS apps cannot scan for Bluetooth devices in the background, limiting the functionality of the finder network. 

This area has gained much new attention since Samsung and Apple have created their Bluetooth item finders. Both companies created enormous finder networks that utilize active smartphones as finder devices~\cite{samsungelectronicsincFindMyMobile2020}.

Several researchers analyzed key finders from manufacturers like Tile, TrackR, and Nut for privacy and security. 
Two new privacy-preserving and end-to-end encrypted key finder protocols have been proposed~\cite{wellerLostFoundStopping2020a, gargSecureCrowdsourcedLocation2021}.

\subsection{The Find My Network}\label{sec:background_findmy}

We summarize the features of the Find My network in this section. We describe its custom \ac{BLE} advertisement format and the working of the device finders available. 
For additional details on the Find My network and its cryptography, we refer to~\cite{heinrichWhoCanFind2021a}. The authors also created OpenHaystack to build custom AirTag-like key finders~\cite{heinrichOpenHaystackFrameworkTracking2021}.
Selected details about the Find My network have also been published by Apple~\cite{appleinc.ApplePlatformSecurity2021}. 

We define the terminology used throughout this paper as follows:
\begin{itemize}
    \item \textit{Finder} devices (e.g., iPhones, iPads, Macs) aid in finding lost or stolen devices.  
    \item The \textit{Find My network} is a network of \textit{finder} devices.
    \item \textit{Find My accessories} are small devices that can be found through Apple's Find My network, including AirTags, AirPods, the Chipolo Spot ONE, and other third-party devices certified by Apple. 
    \item \textit{Find My capable Apple devices} are mainly devices with a screen, including iPhones, iPads, MacBooks, and other Apple devices, which also participate in the Find My network. 
\end{itemize}


\begin{figure}
    \begin{center}
        \includegraphics{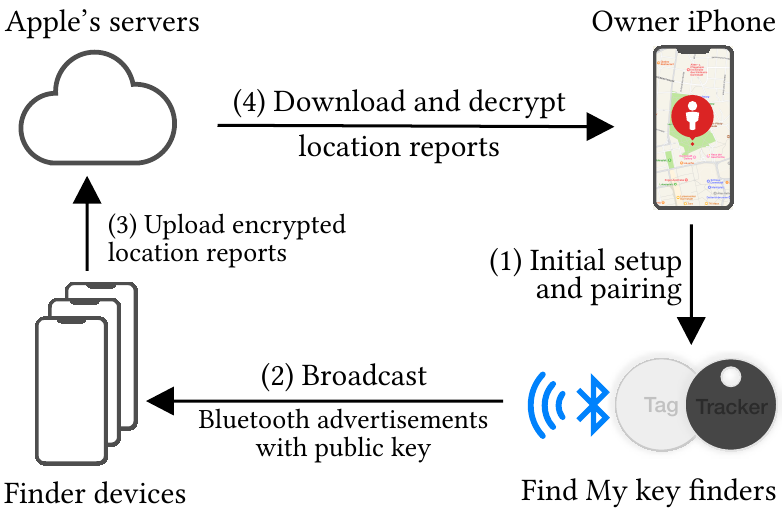}
    \end{center}
    \caption{Simplified Find My network workflow. Adapted from \cite{heinrichWhoCanFind2021a}.}\label{fig:find_my}
\end{figure}

\Cref{fig:find_my} shows a simplified representation of the Find My network workflow. 
The following steps are performed to recover misplaced or stolen devices:
\begin{enumerate*}
    \item The Find My accessories get initialized with an elliptic curve P-224 private-public key pair and a random secret. Those initial keys are called master beacon keys. Beginning with the private-public key pair, the devices can create an infinite number of rotating key pairs by utilizing a key derivation function and the known random secret. 
    \item When the accessory loses its \ac{BLE} connection to the owner's iPhone, it emits the current public key using \ac{BLE} advertisements. For Find My devices: these devices emit their current public key when they lose the internet connection (i.e., a MacBook without WiFi)
    \item Finder devices, which discover such a Find My accessory, extract the public key, generate an ephemeral private-public key pair, and perform a one-sided key exchange using elliptic curve Diffie-Hellman.
    The generated shared secret is then used to encrypt the finder's geolocation. The encrypted location and the finder's public key are uploaded to Apple's servers. 
    \item The owner can now use the Find My application to download and decrypt the location reports. For this, the owner device performs the other side of the key exchange by using the ephemeral public key of the finder and the private key of the Find My accessory, resulting in the same shared secret and allowing the owner to decrypt the location report. 
\end{enumerate*}

For decryption purposes, the master beacon keys of all devices are synchronized in an encrypted form using iCloud. Utilizing the end-to-end encrypted iCloud keychain, any device signed in with the same Apple ID can decrypt the master beacon keys and generate the same private and public keys that were used in the \ac{BLE} advertisements~\cite{heinrichWhoCanFind2021a}. 
So far, it has not been detected that Apple violates the promise not to access the users' private keys~\cite{heinrichWhoCanFind2021a}. Therefore, Apple should not be able to decrypt location reports.

\subsubsection{BLE advertisement format}\label{sec:ble_payload}
A public key on the elliptic curve P-224 can be compressed to $28$ bytes length. 
Apple used some tricks to pack all bytes in one standard \ac{BLE} advertisement while maintaining their common \ac{TLV} based advertisement structure \cite{heinrichWhoCanFind2021a}. 

The first six bytes of the public key form the \ac{BLE} address. Since the address needs to be identified as a static address, the first two bits must be set to \texttt{0b11}. The $22$ remaining bytes are stored in the advertisement's manufacturer data. The manufacturer data follows Apple's standard encoding: The first two bytes are set to the company ID of Apple, the next byte is fixed to \texttt{0x12}, which identifies that this advertisement is used for Find My. The next byte defines the length of the message. Then, the remaining bytes are filled with the public key, a status byte, and a hint byte resulting in the structure shown in \cref{tab:BLE_Payload}. 


\begin{table}
\caption{Find My network advertisement format\\(with zero-index bytes). Adapted from \cite{heinrichWhoCanFind2021a}.}
\label{tab:BLE_Payload}
	\small
	\begin{tabularx}{\linewidth}{rX}
	\toprule
	Bytes & Content \\ 
	\midrule
	0--5 & BLE address ($(p_i[0] \mathbin{|} (0b11 \ll 6)) \mathbin{||} p_i[1..5]$) \\
	\midrule
	6 & Payload length in bytes (\lstinline{30}) \\ 
	7 & Advertisement type (\lstinline{0xFF} for manufacturer-specific data) \\ 
	8--9 & Company ID (\lstinline{0x004C}) \\ 
	10 & Find My network type (\lstinline{0x12}) \\ 
	11 & Data length in bytes (\lstinline{25}) \\ 
	12 & Status (e.g. battery level) \\ 
	13--34 & Public key bytes $p_i[6..27]$ \\
	35 & Public key bits $p_i[0] \gg 6$ \\
	36 & Hint (\lstinline{0x00} on iOS reports) \\
	\bottomrule
	\end{tabularx}
\end{table}

\subsubsection{AirTag and Find My accessories}
Apple opened up the Find My network such that accessories by other manufacturers can be located by it. Those accessories are available in as key finders~\cite{chipolod.o.oChipoloONESpot2021}, bikes, or headphones~\cite{appleinc.AppleFindMy2021}. 
The AirTag was the first device to integrate \ac{UWB} to allow short distance ranging and directional finding.So far, other accessories are not allowed by Apple to implement similar functionalities~\cite{appleinc.FindMyNetwork2020}. Many details about the internals of the Apple AirTag, how to modify the AirTags Firmware, and how to interact with it via Bluetooth have been revealed in \cite{classenHardwearIoNL2021}. 

Find My accessories have four states in which they operate~\cite{appleinc.AppleFindMy2021}: 
\begin{enumerate}
	\item Unpaired 
	\item Connected 
	\item Nearby
	\item Separated 
\end{enumerate}
In the unpaired state, the accessory is waiting to be paired and initialized by an iPhone, iPad, or iPod touch. When the accessory is paired, it stays connected with one of the owner devices (i.e., linked to the same Apple ID). 
If the accessory disconnects from the owner iPhone, it switches into the nearby state. In this state, it advertises the first part of the current public key in the \ac{BLE} address, but it does not advertise the second part in the manufacturer data. Therefore, nearby finders cannot forward the accessory's location. After \SI{15}{\minute}, the accessory transitions from the nearby state into the separated state. Now, it starts advertising the payload described in \cref{sec:ble_payload}, and finder devices can report the accessory's location.
The accessory can always transition back to the connected state from the nearby state and the separated state if an owner device is in proximity.

\subsection{Find My Stalking Protection}\label{sec:background_stalkingProtection}
\subsubsection{Accessories}
All small Find My accessories have to implement simple stalking protections. 
If the accessories stay separated for three days, they start playing a sound when they recognize movement. 
After a sound is played, the accessories will remain silent for at least six hours~\cite{appleinc.FindMyNetwork2020}. 
There are already manuals on the internet showing how an AirTag can be modified to disable the sound feature altogether~\cite{airtagalexHowRemoveSpeaker2021}. In addition, those modified versions are now sold online as well, which lowers the bar for abusive use~\cite{kanSilentAirTagsSpeakers2022}.

\subsubsection{iOS}
Since the release of the AirTags, iOS has contained a tracking detection that identifies when a Find My accessory has followed the user and sends a tracking alert. The user can view a route of the tracking device that has followed him, play a sound on the device, and get information on how to deactivate it.
This system has been evaluated through extensive testing, and several ways to bypass it have been discovered in~\cite{mayberryWhoTracksTrackers2021}. 
We reverse-engineer the implementation in \cref{sec:tracking_detection} and evaluate \AG{} against it in \cref{sec:evaluation}. 

\subsubsection{Android}
In December 2021, Apple released an app called ``Tracker Detect'' on the Google Play Store~\cite{appleinc.TrackerDetectApps2021}. This app was Apple's answer to numerous concerns about location tracking attacks on Android users by AirTags~\cite{ cahnAppleAirTagsAre2021, macAreAppleAirTags2021, koskiDiscoveryAirTagTracking2021}. It allows a user to perform a manual \ac{BLE} scan for Find My accessories. If the app finds such a device nearby, the user has to wait \SI{10}{\minute} before they can play a sound on the device to find it. This restriction is purely software-based, discovered AirTags could play a sound immediately. 
Furthermore, the app contains several guides on how to deactivate known trackers. 

Tracker Detect only offers an insufficient number of features. The app cannot perform automatic background scans, and it cannot warn users that a tracker has followed them for a while. Additionally, self-made Find My tags will not be detected. 

\subsection{Apple Wireless Services}




Researchers have analyzed multiple wireless services used by Apple devices.
AirDrop and Apple's custom WiFi communication layer have been analyzed for security and privacy in~\cite{stuteOneBillionApples2018,stuteBillionOpenInterfaces2019a}.
Many of Apple's wireless services are summarized under the term \textit{Continuity}. Those heavily rely on \ac{BLE} and have been analyzed for privacy issues in~\cite{stuteDisruptingContinuityApple2021,celosiaDiscontinuedPrivacyPersonal2020, martinHandoffAllYour2019}. 
The security of Bluetooth on iOS and the cellular baseband have been analyzed using fuzzing techniques in~\cite{heinzeToothPickerApplePicking2020, krollARIstotelesDissectingApple2021}.
In 2019, \ac{UWB} was added as a new wireless technology in iPhones. The usage of \ac{UWB} in combination with the U1 chip has been analyzed in ~\cite{classenWibblyWobblyTimey2021}. \ac{UWB} provides accurate distance measurements that are secure against relay attacks.  The technology is planned to be used to unlock cars~\cite{appleinc.ExploreUWBbasedCar2021}. Nevertheless, a first attack resulting in a distance reduction has been published~\cite{leuGhostPeakPractical2021a}.


\section{iOS Tracking detection}\label{sec:tracking_detection}
iOS 14.5 added support for the Apple AirTag. This update also included an automatic tracking detection~\cite{appleinc.WhatIfYou2021}. According to experiments by \citeauthor{mayberryWhoTracksTrackers2021}, the tracking detection is triggered by an AirTag that sends \ac{BLE} Find My advertisements while the user is moving, and the AirTag is in proximity to the user's iPhone. When an AirTag or another Find My accessory is marked as ``\emph{suspicious},'' the user will receive a notification. The reception of such a notification can take up to several hours \cite{mayberryWhoTracksTrackers2021}.
We reverse-engineer the tracking detection implemented in iOS and demonstrate our results in this section. Our analysis is based on iOS 15.2, released on December 13, 2021. This detection is available on iPhone, iPad, and iPod Touch. 

\subsection{Methodology}
We use a combined dynamic and static analysis method to reverse-engineer the tracking detection.
All these analyses are performed manually, and we rely on tools to facilitate them. 
\paragraph{Dynamic analysis}
Our goal was to reverse-engineer the current implementation of the automatic tracking detection in iOS. At the time of writing, no jailbreak supports iOS 15.2 or newer. Therefore, our dynamic analysis is based on system logs. 

The tracking detection is executed by the \texttt{locationd} daemon linking the \texttt{TrackingAvoidance} framework. 
Daemons are UI-less processes that the operating system runs to handle tasks, including location processing, Bluetooth, WiFi, and more. Frameworks in iOS are dynamically linked, extending the functionality of processes. 
Both binaries use the system logs to output detailed information: Advertisements received, MAC addresses of Find My accessories, reasons for classifying a device as \emph{suspicious}, and locations where a tracker has been detected. Apple provides several debug profiles to increase the log verbosity~\cite{appleinc.ProfilesLogsBug2022}. We installed the Bluetooth, Location Services, and AirTag profile for this analysis. 
We use the logs to understand the general behavior of the \texttt{TrackingAvoidance} framework and to identify entry points for a static analysis. 

\paragraph{Static analysis}
We base the static analysis on disassembling and decompiling the binary of the \texttt{TrackingAvoidance} framework. 
As the strings in the system logs must be part of the binary, we use those to identify relevant functions in the \texttt{TrackingAvoidance} framework.
The \texttt{locationd} and the \texttt{Track\-ing\-Avoi\-dance} framework are written in Objective-C, which leaves all method names readable for our inspection. 
Our static analysis mainly focuses on the algorithms that classify a Find My accessory as a tracker.

\subsection{System Overview}
\begin{figure}
    \centering
    \includegraphics{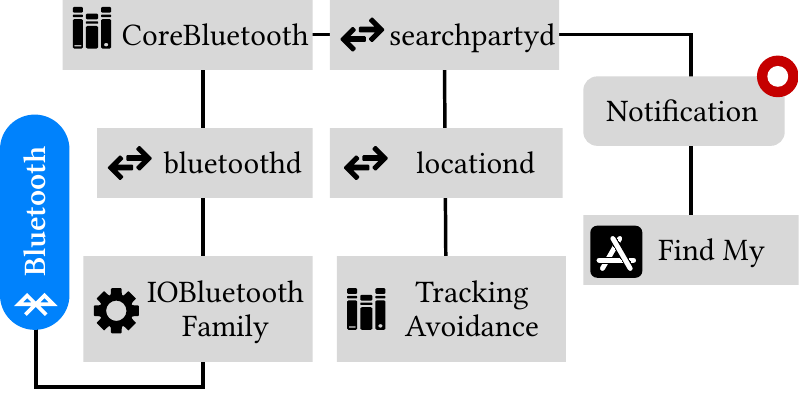}
    \caption{Simplified representation of the iOS tracking detection using components such as apps (\includegraphics[height=0.8em]{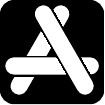}), daemons (\includegraphics[height=0.8em]{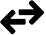}), frameworks (\includegraphics[height=0.8em]{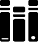}), and drivers (\includegraphics[height=0.8em]{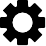})}
    \label{fig:tracking_avoidance_service_map}
\end{figure}

The tracking detection on iOS is a complex combination of several system services that communicate through XPC, Apple's cross-process communication. 
In \cref{fig:tracking_avoidance_service_map}, we highlight the components interacting with each other. 
Many Apple services, e.g., Continuity \cite{martinHandoffAllYour2019, stuteDisruptingContinuityApple2021}, the COVID-19 exposure notification framework~\cite{appleinc.ExposureNotificationApple2020}, and the Find My network require periodic scans for \ac{BLE} advertisements. Therefore, \texttt{bluetoothd} performs these scans and forwards the results to different system daemons. 
All \ac{BLE} advertisements using the Find My format are first forwarded to the \texttt{searchpartyd}. 
This daemon is handling the discovery of lost devices and sends the encrypted location reports to Apple's iCloud servers such that the owner can locate and find their devices~\cite{heinrichWhoCanFind2021a}.
After processing the advertisement, the \texttt{searchpartyd} forwards it to the \texttt{locationd}.
This daemon utilizes the \texttt{TrackingAvoidance} framework, which handles the detection of trackers following the owner of the iPhone/iPad. 
If this framework has classified a device as \emph{suspicious}, it tells the \texttt{searchpartyd} to emit a notification. After opening the notification, the user is forwarded to the Find My app to view a detailed report about the discovered tracker. 

\subsection{Collecting Events for Tracking Detection}

In the \texttt{Track\-ing\-A\-voi\-dance} framework, all Find My advertisements received are checked if they relate to the current user.
All devices logged in with the same Apple ID pre-compute the private and public keys of all devices linked to this account. Those keys are then stored in a read-protected directory on disk for quick look-ups~\cite{heinrichWhoCanFind2021a}. 

Non-owner advertisements are added to a \texttt{TAStore}. 
Besides \ac{BLE} advertisements, the store receives all kinds of events that are partially used to classify if the person is being tracked. These events include regular location updates, user activity events, system state events, and vehicular states. \Cref{tab:tastore_events} shows a list of all known event states.

\begin{table}[]
    \caption{Different event states added to TAStore.}
    \centering
    \begin{tabular}{l l l}
    \toprule
        System States & User Activities & Vehicluar States  \\
        \midrule
        Display on & Static &  Vehicular \\
        DeviceUnlockedSinceBoot & Pedestrian &  Non Vehicluar \\
        HasKoreaCountryCode & Vehicular &  \\
        Wifi & unknown  \\
        LocationServices & & \\
        BatterySaver &  & \\
        HighThermal & & \\
        AP & & \\
        \bottomrule
    \end{tabular}
    \label{tab:tastore_events}
\end{table}

\subsection{Classification of Tracking Devices}
We identified three algorithms implemented by the \texttt{Track\-ing\-A\-voi\-dance} used to identify \emph{suspicious} devices. 
The first one is the \textit{general detection}, which is executed regularly and iterates over all devices saved in the \texttt{TAStore}. 
Significant location changes, so-called visits, set off the other two.
We identified the default values for many parameters in the framework. However, the actual values may change during execution, and as they are not logged, we cannot inspect them. 

\subsubsection{General detection}\label{sec:trackingAvoidance_generalFilter}
Every two to five minutes, the \texttt{locationd} runs a classification on all devices from which advertisements have been received in a recent time frame (default: \SI{15}{\minute}).  
The algorithm defines a threshold duration (default: \SI{10}{\minute}) and a threshold distance (default: \SI{840}{\metre}) for which the devices had to follow the person. 

The classification of all devices follows the pseudo-code shown in \cref{alg:classification}.
Essentially, it checks if any device exceeds both thresholds, if the recent user activity is somehow plausible and if, additionally, a people density scan measurement is used. We could not find clear indications if the people density scan was performed during our dynamic analysis. In iOS 15.2, the people density scan does not seem to be activated. 
For every \emph{suspicious} device, iOS sends a notification to the user. 

\begin{algorithm}
\caption{General Filter that classifies suspicious devices.}
\label{alg:classification}
\begin{algorithmic}
    \State $devices \gets TAStore.devices$
    \For{$ device~in~devices$}
        \State $isInVehicle \gets TAStore.lastVehicularState.isInVehicle$
        \State $peopleDensity \gets TAStore.lastPeopleDensity$
        \State $userActivity \gets TAStore.dominantUserActivity$
        \State $walkingSpeed \gets TAStore.walkingSpeed$
        
        \State $dist \gets device.distanceTravelled$
        \State $dur \gets device.durationTravelled$
        
        \State $isDriving \gets isInVehicle \mathbin{|} peopleDensity$
        \State $isDriving \gets isDriving \mathbin{\&} (userActivity = vehicular)$
        \\
        \State $isWalking \gets (userActivity = pedestrian)$
        \State $isWalking \gets isWalking  \mathbin{\&} (walkingSpeed < MAX\_SPEED) $
        
        \State $validMovement \gets isDriving \mathbin{|} isWalking $
        
        \State $minDistTravelled \gets dist > THRESHOLD\_DISTANCE$
        \State $minDurTravelled \gets dur > THRESHOLD\_DURATION$
        
        \State $result \gets valid Movement$
        \State $result \gets result \mathbin{\&} minDistTravelled$
        \State $result \gets result \mathbin{\&} minDurTravelled$
        
        \If{$result=1$}
        \State mark~$device$~as~$suspicious$
        \EndIf
        
    \EndFor
\end{algorithmic}
\end{algorithm}

\subsubsection{Visit-based detection}
The public \texttt{CoreLocation} framework implements the \texttt{CLVisit} objects, informing an app that a person is currently ``visiting'' a specific location. Those objects contain an arrival and a departure time, and they are generated when the user arrives or leaves a location. 
Apple does not specify the duration a person needs to be at a location to generate a \texttt{CLVisit} \cite{appleinc.CLVisitAppleDeveloper2014}. 
Based on these objects, the \texttt{TrackingAvoidance} framework observes user visits and combines this information with \ac{BLE} advertisements saved to the \texttt{TAStore}. 

The first algorithm, \texttt{TADetectionTypeVisit} is simple. It creates an intersection between the devices seen at the last visit and the current visit. All devices in this intersection are marked as \emph{suspicious} devices. No other checks are performed. The distance and the duration that the device has traveled with the user are stored. 

The second algorithm, \texttt{TADetectionTypeSingleVisit} uses the intersection created before and performs extra checks on the devices. 
If all checks are evaluate to \textit{true}, the device will be marked as a \emph{suspicious} device. 
For each device, the algorithm checks:
\begin{enumerate}
    \item Has the last advertisement been received recently (default: in the last \SI{5}{\minute})?
    \item Is the device traveling with the user for a longer distance than the threshold distance (default: \SI{420}{\metre})?
    \item Is the device traveling with the user longer than the threshold duration (default: \SI{300}{\second})?
\end{enumerate}

The dynamic analysis has shown that both algorithms are in active use to classify a device as a \emph{suspicious} device. Interestingly, the threshold distance in the \texttt{TADetectionTypeSingleVisit} uses only half the distance than the general detection.


\subsection{Tracking Device Categories}
In iOS 15.2, the system identifies four different device types named \texttt{Other}, \texttt{D}, \texttt{H}, and \texttt{HELE}. 
Bit 2 and 3 of the status byte in the BLE advertisement categorize the devices.
We identified categories for each type and listed them in \cref{tab:device_types}. 

\begin{table}[]
    \caption{Different device types in iOS TrackingAvoidance} \label{tab:device_types}
    \centering
    \begin{tabularx}{\linewidth}{l l X X}
    \toprule
        Device Type & Bits & Category & Example  \\
        \midrule
        Other & $0b00$ & Apple devices & iPhone, Mac, iPads  \\
        D (Durian) & $0b01$ & AirTags & AirTag \\
        H (Hawkeye) & $0b10$ & 3rd Party & Chipolo ONE Spot \\
        HELE & $0b11$ & Headphones & AirPods Pro \\
        
        \bottomrule
    \end{tabularx}
\end{table}


Devices that are marked as ``Other'' are, in most cases, Apple devices such as MacBooks that lost their internet connection and are now sending Find My advertisements~\cite{heinrichWhoCanFind2021a}. Apple sets the status byte to \texttt{0x00}. Note that such devices are marked as ``irrelevant'' for tracking in the system logs. 
Previously \citeauthor{heinrichOpenHaystackFrameworkTracking2021} created a tag that copied the behavior of a MacBook with Find My, which essentially creates a tracker that evades Apple's tracking detection, an issue initially discovered and reported by \citeauthor{mayberryWhoTracksTrackers2021}~\cite{heinrichOpenHaystackFrameworkTracking2021,  mayberryWhoTracksTrackers2021}.
Using our reverse-engineering approach, we can confirm that this issue persists in iOS 15.2. 


\subsection{Notifications}
At what time a notification is delivered is defined by several factors. 
When the user has returned home and and the \emph{suspicious} devices continue to appear during the \ac{BLE} scans, the system sends out a notification promptly.
If the user is not returning home, the notifications are delayed. In this case, the \texttt{Track\-ing\-Avoi\-dance} marks \emph{suspicious} devices as ``staging.'' Every staging phase contains an end date. 
After the staging has ended, the \texttt{locationd} can either send a notification immediately or prolong the staging phase.
We could not identify the trigger that determines when a notification is sent if the user does not return home. The evaluation in \cref{sec:evaluation_results} shows that it can take several hours. 

\subsection{Discussion}
Apple's \texttt{TrackingAvoidance} framework implements a tracking detection that leverages the information from many different sources and frequent (every \SI{1}- \SI{5}{\minute}) \ac{BLE} scan intervals.
Many of those features are not available to third-party applications. The downside of this framework is that Apple neglects the fact that any \ac{BLE} capable device can be transformed into a Find My network tracker by mimicking the behavior of an iPhone as demonstrated by the OpenHaystack framework~\cite{heinrichOpenHaystackFrameworkTracking2021}. Users will not get a notification for those trackers.

Apple's tracking detection identifies trackers early on. However, it also delays the delivery of notifications.  
The only way to reliably get a notification is by returning to the home location, which essentially gives the home location away to the attacker. 
A possible reason why notifications are delayed is that Apple wanted to minimize the risk of false positives. 

At last, Apple's Bluetooth API disallows access to Bluetooth scans in the background. Removing those limitations could allow third-party developers to create apps that detect trackers from other manufacturers. 




\section{AirGuard}\label{sec:airguard}

After the AirTag release, Apple left Android users without tracking protection for over six months until December 2021. The design and implementation of \AG{} started in April 2021, and the first public version was released in August. 
The automatic tracking protection is intended to work similarly to the one implemented in iOS.
Using regular \ac{BLE} background scans, \AG{} detects all kinds of Find My capable tracking devices, including self-made ones. In less than \SI{45}{\minute} users receive a notification about the tracker and guidance on finding it. 
In this section, we introduce \AG{} and its tracking detection algorithm. We then explain how we designed the user interface to cater to non-expert users. 

\subsection{Tracking Detection}\label{sec:airguard_trackingDetection}

The tracking detection of \AG{} is a combination of timed \ac{BLE} background scans and a detection algorithm that classifies malicious devices as location trackers. 
\subsubsection{Bluetooth scanning}
The app uses the Android Work Manager to schedule recurring background tasks to perform the tracking detection.
In these tasks, the app performs a \ac{BLE} scan and runs our tracking detection. In Android, such tasks can be executed at maximum every \SI{15}{\minute}, which limits the time to detection.  

During a scan, the app filters the results for Find My advertisements in the separated state (see \cref{sec:background_findmy}).
\AG{} fetches the current geolocation and stores all advertisements received together with the location in a local database.
When Find My accessories are in the nearby state, they cannot be located by finders, and therefore \AG{} ignores them. 

To achieve high reliability on detecting trackers, the scan duration is set to \SI{8}{\second} and the scan Android scan mode to low latency. More details on these parameters are listed in \cref{apdx:scan_parameters}.

\subsubsection{Tracker classification algorithm}
\begin{figure*}[t!]
    \centering

    \begin{subfigure}[t]{0.2\textwidth}
        \centering
        \includegraphics[width=\textwidth]{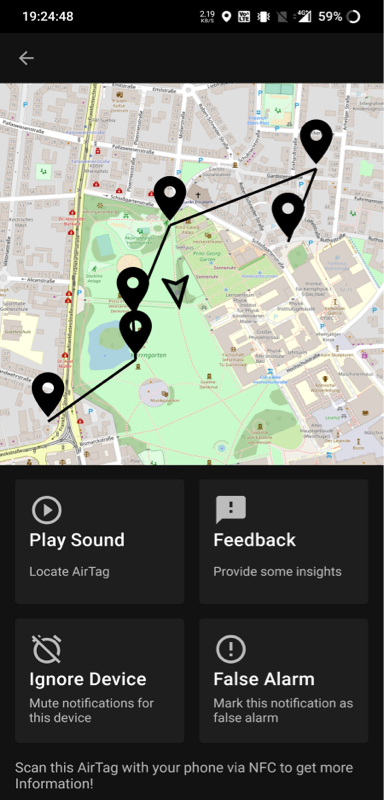}
        \caption{Information after a tracking alert.}
        \label{fig:tracking_alert_detail}
    \end{subfigure}
         \hfill
    \begin{subfigure}[t]{0.2\textwidth}
        \centering
        \includegraphics[width=\textwidth]{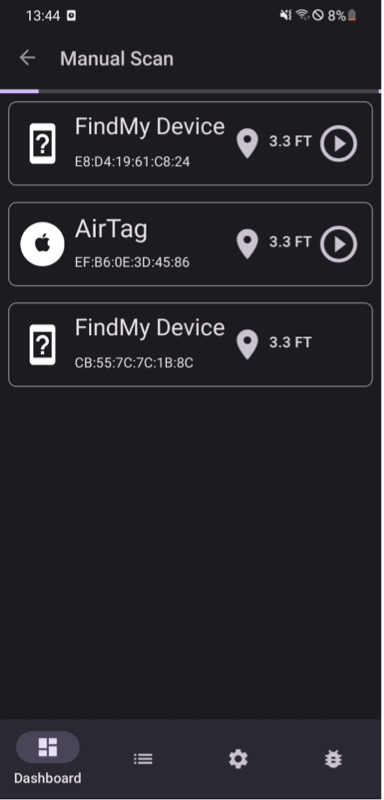}
        \caption{User performing a manual scan to find trackers.}
        \label{fig:manual_scan}
    \end{subfigure}
        \hfill
    \begin{subfigure}[t]{0.2\textwidth}
        \centering
        \includegraphics[width=\textwidth]{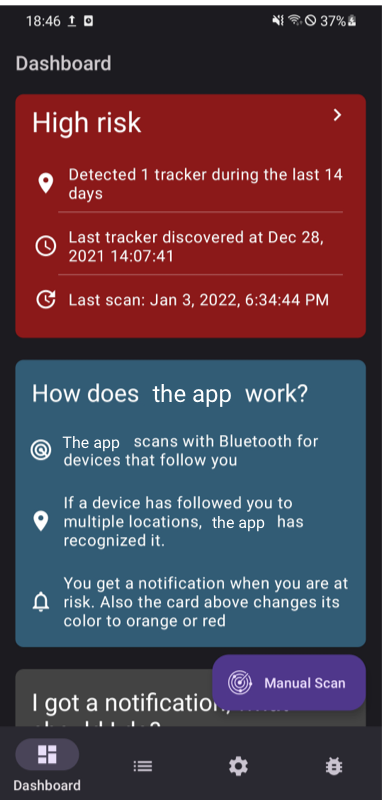}
        \caption{User with a high risk of being tracked.}
        \label{fig:dashboard}
    \end{subfigure}
         \hfill
    \begin{subfigure}[t]{0.2\textwidth}
        \centering
        \includegraphics[width=\textwidth]{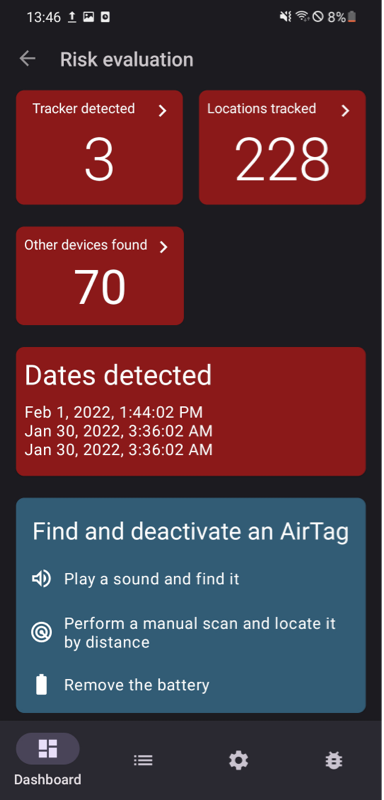}
        \caption{Details on how trackers follow the user.}
        \label{fig:risk_score_detail}
    \end{subfigure}
    
        \caption{Screenshots of \AG{}.}
    \label{fig:screenshots}
\end{figure*}

After every scan, the app runs our detection algorithm, which iterates over all devices in the database. 
Generally, \AG{} identifies different devices based on their \ac{BLE} MAC address. As the MAC address is formed by using the current public key, it does not change until the public key changes. 

For every device, the algorithm performs the following checks: 
\begin{enumerate}
    \item Have the device's advertisements been received for longer than the minimum duration of \SI{30}{\minute}. 
    \item Has the device been discovered at least three times.
    \item Has the device traveled with the user for a minimum distance of \SI{400}{\meter}.
    \item Has there already been a tracking alert in the last seven hours? \AG{} limits the number of notifications to one per device every for this duration. 
    \item If all checks evaluate to \texttt{true}, a notification will be delivered to the user immediately. 
\end{enumerate}

In general, our algorithm is a simplified version of Apple's general tracking detection as described in \cref{sec:trackingAvoidance_generalFilter}. 
We use a minimum distance and a minimum duration to limit the number of false positives when warning a user. 
Therefore, an AirTag of a neighbor or a family member does not result in a notification.

Since we identify devices based on their \ac{BLE} MAC address, they will be recognized as a new device when the public key changes. Our main targets are Apple AirTags, and other Find My accessories.  According to the specification, Find My accessories update their public key once a day around 4 a.m. local time~\cite{appleinc.FindMyNetwork2020}. 

\subsubsection{Notifications}
\AG{} delivers tracking alerts as Android notifications. In the current implementation, the notifications are delivered early. 
Early notifications should help users to identify a potential tracker quickly and even before they return home and might reveal their home location to someone else. 

When users open the notification, they get a detailed overview of the locations where the tracking device has been detected. \Cref{fig:tracking_alert_detail} shows a screenshot of the user interface presented to the user when a tracking notification is opened. 
Here, the user can play a sound on the device, ignore it, send feedback, and mark this notification as a false alarm. The feedback is basic information on where the tracking device has been hidden (car, bike, backpack, clothes), and it will be sent to the developers if the user has agreed to share data. 

\subsection{Manual Scanning}
Finding a tracking device can be difficult, especially if they are self-made or modified such that the AirTags cannot play a sound~\cite{kanSilentAirTagsSpeakers2022}. In this case, a user can manually initiate a scan process and see an automatically updating estimated distance of all potential tracking devices in the vicinity. \Cref{fig:manual_scan} shows a screenshot of this process.
\AG{} estimates a distance based on the \ac{RSSI} and sample measurements with an AirTag. Precision varies due to the design of Bluetooth. 
If the device does support playing a sound, this can also be performed from here without any artificial waiting period. 

\subsection{User Experience}\label{sec:airguard_riskLevel}
The app's main view is a dashboard, as shown in \cref{fig:dashboard}. 
This dashboard should allow non-experts to identify their personal risk at one glance. 
The main element is the risk level card, which represents the user's risk of being tracked in three colors:
\begin{enumerate}
    \item Green (Low): no tracking device following the user has been detected in the last 14 days.
    \item Orange (Medium): one or multiple devices have been detected that followed the user for less than 24 hours.
    \item Red (High): one device or multiple devices have been detected that followed the user for more than 24 hours.
\end{enumerate}
Additional details on how many trackers were found and when the last one has been detected are displayed as well. 
When the user taps on the card, they get more details (see \cref{fig:risk_score_detail}): the number of locations that have been recorded with the tracker and the number of Find My devices that have been found in total. 

From here on, users can get more information by viewing a list of all devices that have been marked as tracking the users' location. They can view all locations where the device has been detected on a map and initiate the sound on Find My accessories.
At last, users can also view all Find My devices found by the app even if they have not followed them. Expert users who want to validate the tracking detection behind \AG{} highly requested this feature.




\section{Evaluation}\label{sec:evaluation} 

This section evaluates and compares \AG{} to the iOS tracking detection by testing them in three scenarios.
The focus is set on: devices that can be detected, placements of the devices, the time it takes until the user is notified, and the number of locations that have been until a notification was sent. 

\paragraph{Attacker model}
We define our attackers as people with basic technical knowledge who have access to an Apple device and buy an off-the-shelf tracker by Apple or a Find My certified manufacturer. We include attackers who can use the manuals of OpenHaystack to create a self-made tracking device. Using this device, they try to track the location of their victim. 
Advanced attackers with large resources who can modify the firmware of AirTags or write custom firmware for self-made tags are out of scope. Those attackers already have access to many different tracking technologies like GPS-based trackers.  

\subsection{Setup of the Experiments}
To evaluate our app and the iOS tracking detection, we use the following devices: for the iOS tracking detection, we use an iPhone 12 mini that runs iOS 15.2. Find My and the Find My Network are activated on the iPhone. 
For the \AG{} tracking detection, we use a Samsung S21 Ultra 5G.
\AG{} uses the following settings: location access in the background is granted, battery optimization is disabled. Before each test, we clear the database of \AG{}.

We test the app against three different devices: an Apple Air\-Tag~\cite{appleinc.AirTag2021}, a Chipolo ONE Spot~\cite{chipolod.o.oChipoloONESpot2021}, and a self-made OpenHaystack tag~\cite{heinrichOpenHaystackFrameworkTracking2021}. 
The AirTag and the Chipolo tag follow Apple's reference implementation of the Find My accessories~\cite{appleinc.AppleFindMy2021}. 
During the evaluation experiments, all tags were kept in a separate state.
The self-made OpenHaystack tags essentially copy the \ac{BLE} messages of a lost iPhone. They keep a static public key and do not update it. 
We use the following placements for our testing: In a pocket of a victim's coat, inside a backpack, and attached to a car. 
The first two placements are ideal for tracking a person's movement. The currently available devices can be easily hidden inside a backpack, making it difficult for a victim to discover them. 
Tagging vehicles has been chosen because of recent media coverage of thieves attaching AirTags to high-end cars to track vehicles back to their parking location~\cite{koskiDiscoveryAirTagTracking2021,macAreAppleAirTags2021}.

\subsection{Results}\label{sec:evaluation_results}
We divide the results of our evaluation into sections based on the placements of the trackers and summarize the results in \cref{tab:evaluation_results}.
We utilize all three trackers in all scenarios to identify how the different systems recognize different tracking devices.

\paragraph{Pocket} For the trackers' pocket position we were able to receive tracking notifications in \SI{35} minutes with \AG, which is much faster than it took iOS to send a notification. Even though iOS did recognize the tracking devices early on and continued to scan for them in frequent intervals, the system decided to delay the tracking notification. In this evaluation the notification was sent exactly when we returned to the iPhone's \textit{home} location. 

\paragraph{Backpack}
For the backpack position of the trackers, \AG{} took \SI{31} minutes to locate the trackers and send a notification. 
For iOS, it took more than \SI{4} hours to send a notification. 
Both devices were recognized early on and linked to over 70 locations.
Again, the main reason for iOS to delay notifications is the user's current location. We did stay away from the home location for \SI{6} hours in this scenario. 

\paragraph{Car}
News reports mention that the fuel cap is a commonly used position by attackers who do not have access to the inside of the car~\cite{charltonAppleAirTagItem2021}. We glued an AirTag and a Chipolo ONE Spot to the inside of the fuel cap. For space reasons, we placed the OpenHaystack tag inside the car. We took two \SI{30} minute rides with the prepared car and stayed away from the car for \SI{3} hours and \SI{30} minutes in between. \AG{} sent a tracking notification during the second ride, when the trackers are detected for the third time. The iOS tracking detection never sent a notification, not even after returning to the iPhone's home location.

\begin{table}
\caption{Tracking Detection Results}\label{tab:trackin_ios}
\label{tab:evaluation_results}
	\small
	\begin{tabularx}{\linewidth}{lXXX}
    \toprule 
    Tracker position & Tracker & Time to notification & Locations with tracker until notification \\
    \midrule
    \textbf{iOS} & & & \\
    Pocket & AirTag & 1h 45m 14s & 35 \\
    Pocket & Chipolo & 1h 45m 14s & 28 \\
    Pocket & OpenHaystack & - & - \\
    \specialrule{0.25pt}{1pt}{1pt}
    Backpack & AirTag & 4h 14m 23s & 87 \\
    Backpack & Chipolo & 4h 14m 23s & 74 \\
    Backpack & OpenHaystack & - & - \\
    \specialrule{0.25pt}{1pt}{1pt}
    Car & AirTag & - & - \\
    Car & Chipolo & - & - \\
    Car & OpenHaystack & -  & - \\
    \midrule
    \textbf{\AG} & & & \\
    Pocket & AirTag & 35m 20s & 3 \\
    Pocket & Chipolo & 35m 20s & 3 \\
    Pocket & OpenHaystack & 35m 20s & 3 \\
    \specialrule{0.25pt}{1pt}{1pt}
    Backpack & AirTag & 30m 49s & 3 \\
    Backpack & Chipolo & 30m 49s & 3 \\
    Backpack & OpenHaystack & 30m 49s & 3 \\
    \specialrule{0.25pt}{1pt}{1pt}
    Car & AirTag & 4h 18m 11s & 3 \\
    Car & Chipolo & 4h 18m 11s & 3 \\
    Car & OpenHaystack & 4h 18m 11s & 3 \\
    \bottomrule 
	\end{tabularx}
\end{table}

\subsection{Discussion}\label{sec:evaluation_discussion}

\TODO{Section in present tense}
Our evaluation shows that \AG{} detected all three trackers in all scenarios. 
The app uses an aggressive detection algorithm, which resulted in a notification after three discoveries for all scenarios. 

\paragraph{Car}
Unfortunately, the iOS tracking detection failed to detect trackers that were not permanently close to the user in our car scenario. 
Based on our results in \cref{sec:tracking_detection}, we try to identify the reason for these missed notifications.
The general detection algorithm only uses the advertisements received in the last \SI{15}{\minute}.
If the user now stays out of the vicinity for more than \SI{15}{\minute}, the trackers will be forgotten by iOS. However, the threshold duration of \SI{10}{\minute} and the threshold distance of \SI{840}{\meter} have been exceeded in our test twice.
We expect the default values to change during execution and based on user behavior.
If this has been the case, it would explain why no notification has been delivered. Media reports have mentioned that people have been notified of AirTags on the car~\cite{macAreAppleAirTags2021}. 
A straightforward mitigation would be to integrate all received advertisements of the current day in the tracking detection. 

\paragraph{Notifications}
The iOS' tracking detection delivers notifications much later, depending on the user's location. Apple has announced to mitigate this issue in an update~\cite{kirschnerUpdateAirTagUnwanted2022}. 
Furthermore, the evaluation has confirmed that self-made tags with OpenHaystack will not be detected~\cite{mayberryWhoTracksTrackers2021}. 

\paragraph{Scan opportunities}
One weakness of \AG{} is the limited scan opportunities the Android operating system grants it. Therefore, \AG{} finds fewer locations with a tracker. 
As a comparison, iOS detected the AirTag at 35 locations in \SI{1}{\hour} \SI{45}{\minute} in the pocket scenario and \AG{} detected it at seven locations. 

\paragraph{False positives}
The limited scan opportunities and the intentional early notifications can also lead to false positives. During our evaluation, we did not experience any false notification, but we can imagine several scenarios that would trigger one: 
\begin{enumerate*}
\item Travelling in a train or airplane next to an AirTag in separated state (e.g., due to Airplane mode) \AG{} will issue a notification. 
\item GPS drifts. When using \AG{} indoors and the smartphone fails to get a good GPS signal or fails to receive a location update based on WiFi signals, the location might drift away for more than our minimum distance. If an AirTag is close in this case, it will appear to be following the user, and the app may send a notification.
\end{enumerate*}

Those false positives may not be harmful to most users because the user can then match the reported locations to a specific event based on the provided map. 
However, the risk score of the app will increase to medium if the user receives only one false positive in 14 days. 
We keep possible solutions for this for future work. 

\paragraph{Fast rotating trackers}
\AG{} and the iOS tracking detection, do not detect specially modified tracking devices, which rotate their public key more than once an hour~\cite{mayberryWhoTracksTrackers2021}. As the public keys are unlikable, the device would be detected as a new device after a key rotation. Attackers require advanced technical knowledge to deploy such an attack and cannot buy such devices off the shelf.


\section{User study}\label{sec:user_study}

    
Since the first release of \AG{}, users have been able to share anonymized data. The shared data allows identifying potential problems of the app and getting hints on the prevalence of tracking attacks in the wild. 

\paragraph{Provided data}
We ask all users if they are willing to share anonymous data with us during the app setup. If they agree, the app will regularly upload received advertisements, devices, tracking notifications sent, and optional user feedback.
Before the data is sent, all personally identifiable information is removed, such as the \ac{BLE} MAC address and the \ac{BLE} payload. Therefore, if two different users receive the same advertisement, we store it as two different advertisements.  
With over $120\,000$ active users, of which over $30\,000$ donated anonymized data, we created a dataset with more than $8\,600\,000$ Find My advertisements received. 

\paragraph{Limitation}
We limit the evaluation of our data to a six-week time frame starting on December 15, 2021, and ending on January 26, 2022, as we changed our data collection framework with an update on December 11, 2021.


\subsection{Discovered Devices}\label{sec:userStudy_discoveredDevices}

\begin{figure}
    \begin{center}
        \includegraphics{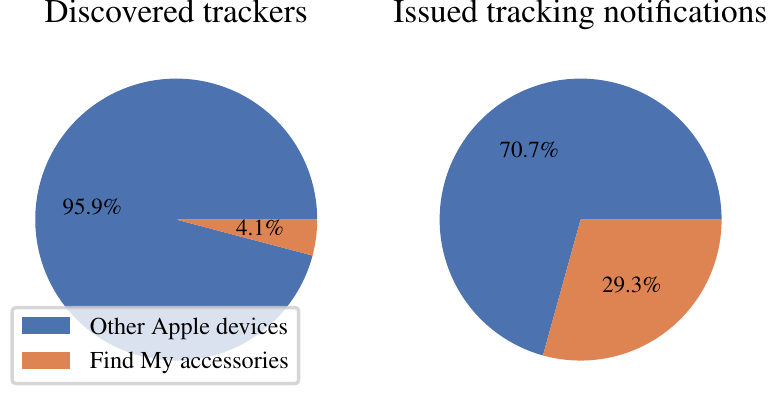}
    \end{center}
    \caption{Device type distribution for discovered trackers as well as issued tracking notifications.}\label{fig:device_distribution}
\end{figure}

\begin{figure}
    \begin{center}
        \includegraphics{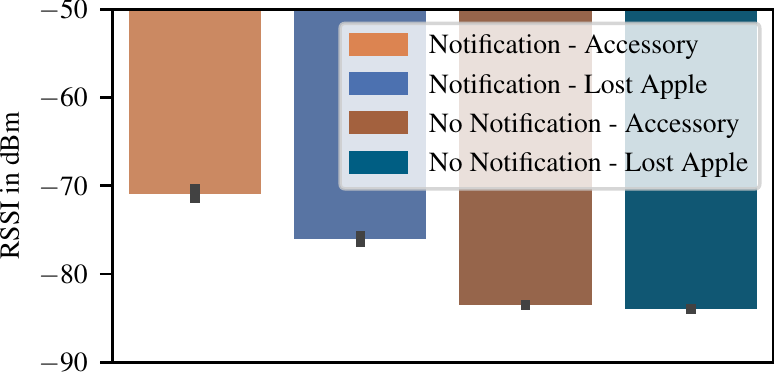}
    \end{center}
    \caption{Mean RSSI comparing tracking devices with normal devices.}\label{fig:rssi_comparison}
\end{figure}

We can separate the devices that \AG{} found into two groups: Find My accessories (e.g., AirTags) and Find My capable Apple devices (e.g., MacBooks) as defined in \cref{sec:background_findmy}.
Self-made tags copy the behavior of Apple devices and are therefore assigned to the same group~\cite{heinrichOpenHaystackFrameworkTracking2021}. 
In total, users have discovered over $3$ million devices that use Find My in our defined time frame. Only about $160\,000$ were Find My accessories. All devices together caused $2494$ notifications for $1325$ users. 
As shown in \cref{fig:device_distribution}, $29\%$ of all notifications were caused by Find My accessories, while they represent only $4\%$ of all devices found. 

\Cref{fig:rssi_comparison} shows the mean \ac{RSSI} based on the device type and if the device issued a tracking notification. 
The app captures the \ac{RSSI} for all advertisements received by \AG{} users. It hints at how close a sending device is to the user during the reception, but it does not allow precise distance measurements. 
The dataset has two takeaways: 
\begin{enumerate*}
    \item Devices that trigger a notification are closer to the user.
    \item Accessories are closer to the user than Apple devices.
\end{enumerate*}


\paragraph{Limitations}
From the current dataset, it is impossible to state if notifications have been initiated on purpose by users who wanted to test \AG{}, how many of the notifications were caused by self-made tags, and how many Apple devices were falsely classified as malicious trackers. 
Although self-made tags offer a seven-day location history not available in the Find My app, they are less accessible and require a manual setup.  
This implies that most Apple devices have been false positives because using an iPhone or a MacBook to track someone's location is impractical. 
Moreover, users can report a false alarm actively, and during our evaluation, $97$ notifications were marked as false alarms, and $85$ of those were caused by Find My capable Apple devices

As \AG{} needs to recognize a device for at least \SI{30}{\minute}, these notifications signify that not all Apple devices stick to the \SI{15}{\min} public key update interval identified in~\cite{heinrichWhoCanFind2021a}.
If Apple devices do not permanently keep that interval, it would also explain why Apple refrained from warning users if such a device follows them.
A simple update to limit the number of false positives in \AG{} and integrate this group of devices in iOS would be to require a longer minimum duration for lost Apple devices to cause a notification.

\subsection{Location Tracking Prevalence}

\begin{figure*}[t!]
    \centering
    
    \begin{subfigure}[b]{0.47\textwidth}
        \centering
        \includegraphics[width=\textwidth]{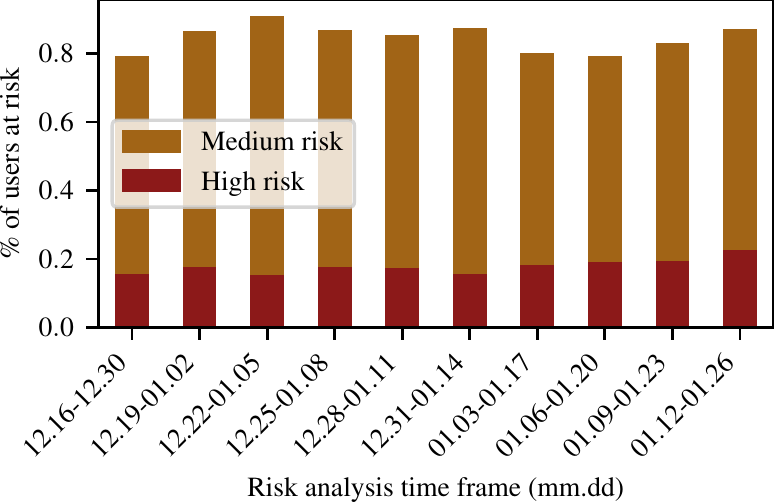}
        \caption{Percentage of data donors at a medium and high risk levels from December 15, 2021 to January 26, 2022.}
        \label{fig:risk_level_normalized}
    \end{subfigure}
    \hfill
    \begin{subfigure}[b]{0.47\textwidth}
        \centering
        \includegraphics[width=\textwidth]{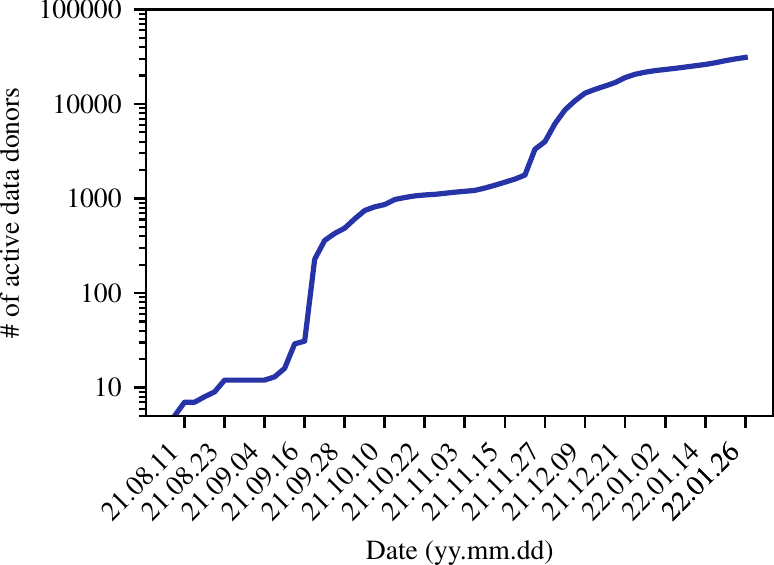}
        \caption{Overall active data donors since the release of the app.}
        \label{fig:user_statistics}
    \end{subfigure}

    \caption{Statistics from data donors}
    
\end{figure*}

The main goal of \AG{} is to serve as a tool that protects its users from abuse through stalking by using an AirTag or a similar Find My accessory. This section analyzes the data to discover hints on the prevalence of location tracking in the wild. 
We limit the data to only Find My accessories, such as AirTags and the Chipolo ONE Spot. We do not cover self-made trackers and lost Apple devices because they can cause more false positives (see \cref{sec:userStudy_discoveredDevices}).  

As described in \cref{sec:airguard_riskLevel}, the app calculates a user-based risk level which is based on the time a user has been tracked. The risk level always accounts for trackers detected in the last two weeks, and it can be low, medium, or high. 

\Cref{fig:risk_level_normalized} shows the percentage of data donors marked with a medium or a high-risk level. 
The number of data donors in a specific risk level results from calculating every user's risk level at the time frames shown on the horizontal axis using a two-week sliding window. We then calculate the percentage by dividing through all active data donors.
An active data donor is defined as a user who has submitted any data during the two-week time frame. It shows that around $0.2\%$ of data donors are constantly ranked at a high risk of being tracked. 
The percentage for medium-risk data donors varies from $0.6\%$ to $0.8\%$. Interestingly, the fraction of users at risk stays rather constant, even with a growing user base.

\Cref{fig:user_statistics} shows the number of active data donors since the release of \AG{}. The first steep increase of new data donors is related to a hacker news post. After a woman who has been a victim of a tracking attack mentioned our app in a TikTok video\footnote{\url{https://www.tiktok.com/@angel.edge95/video/7035276471237217541}} \AG{} received a second steep increase.

\begin{figure}
    \begin{center}
        \includegraphics{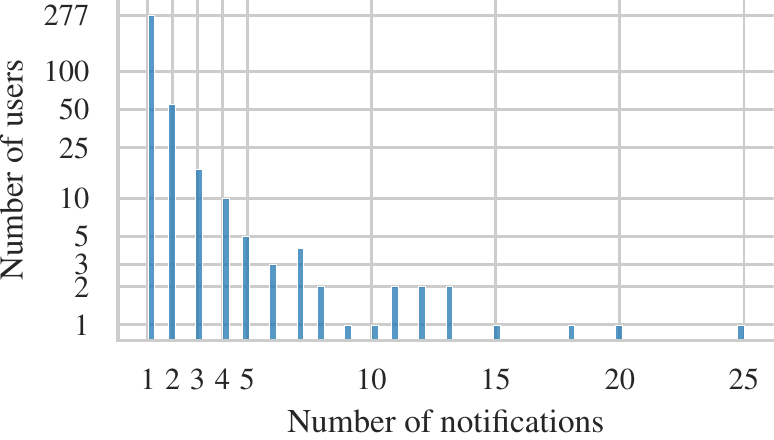}
    \end{center}
    \caption{Density distribution over number of notifications received per user.}\label{fig:notification_distribution}
\end{figure}

\Cref{fig:notification_distribution} shows a histogram over the number of notifications a user has received. 
Most users receive only one notification, while some individuals received up to $25$ notifications in our six-week observation period. 






\subsubsection{User feedback}
Users can give feedback on identified trackers. This can be that a tracker is marked as a false alarm or that the user provides us with basic information where a tracker has been hidden. 
During our evaluation, only five users reported the location of identified trackers: one user declared that the tracking device was placed in their backpack, and four others reported a tracker in or on their car. 
Besides our feedback through the app, we also received reports from multiple users through reviews on the Google Play Store or email. One user who wants to stay anonymous messaged us via email: ``\emph{I have found 2 Airtags attached to my car \dots I don't really know what else to do or say.  I just wanted to say thank you for your app.}'' Several screenshots of the app showing tracking devices found have been attached to this email. 





\section{Conclusions}\label{sec:conclusion}
We reverse-engineered the iOS tracking detection to analyze its limits.
Furthermore, we confirmed a way to bypass the detection using self-made trackers in iOS 15.2, initially discovered in iOS 14.5. 
We designed, implemented, and released the Android app \AG{}, an open-source anti-tracking solution against Find My accessories. $120\,000$ active users show a high demand for automatic tracking detection. 
The evaluation showed that \AG{} found more actual trackers in different scenarios compared to the iOS tracking detection. 
We analyzed a dataset generated by over $30\,000$ \AG{} users. In general, the dataset showed that the tracking detection is triggered in practice and several user reports confirmed. 
We hope our work leads to better tracking protection mechanisms implemented by tracker manufacturers directly. 

\bibliographystyle{ACM-Reference-Format}
\bibliography{AirGuard}


\appendix 

\section{\AG{} - Additional details}

\subsection{BLE Scan Parameters}\label{apdx:scan_parameters}
\begin{figure}[b]
    \centering
    \includegraphics{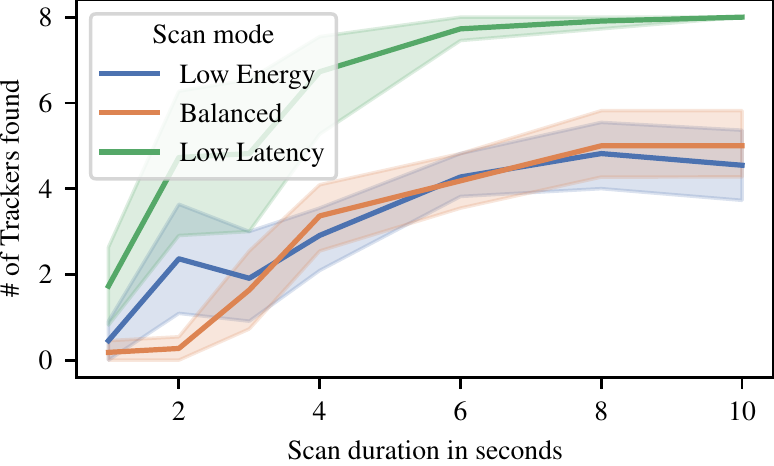}
    \caption{Evaluation of different scan parameters}
    \label{fig:scan_evaluation}
\end{figure}

Android's \ac{BLE} scans use two parameters that can enhance the number of scan results or save energy: The scan duration and the scan mode. 
To optimize our systems configuration, we tested all combinations of those parameters with ten consecutive scans. We tested a scan duration from \SI{1}{\second} to \SI{10}{\second} against all available scan modes. 
\Cref{fig:scan_evaluation} shows the results of this evaluation with eight tracking devices in range.

\begin{table*}[ht]
\caption{Characteristics to start and stop playing a sound on Find My accessories}
\label{tab:characteristics}
	\small
	\begin{tabularx}{\linewidth}{lllll}
    \toprule
    Device & Behavior & Service & Characteristic & Value to write \\
    \midrule
	AirTag & Play sound & \texttt{7DFC9000-7D1C-4951-86AA-8D9728F8D66C} & \texttt{7DFC9001-7D1C-4951-86AA-8D9728F8D66} & \texttt{0xAF} \\
	Accessories \& AirPods & Play sound & \texttt{FD44} & \texttt{4F860003-943B-49EF-BED4-2F730304427A} & \texttt{0x01 00 03} \\
	Accessories \& AirPods & Stop sound & \texttt{FD44} & \texttt{4F860003-943B-49EF-BED4-2F730304427A} & \texttt{0x01 01 03} \\
	\bottomrule
	\end{tabularx}
\end{table*}

\section{Sound on Accessories}\label{apdx:sound_accessories}

Whenever a tracker has been detected by the iOS tracking detection or the Android App Tracker Detect, the user can play a sound on the accessory. Playing a sound on a tracker significantly simplifies the discovery of the device. The same functionality has been adopted in \AG{}. 

Apple has not documented how this action can be triggered. The Find My network accessory specification defined a \ac{BLE} \ac{GATT} protocol to play and stop a sound~\cite{appleinc.FindMyNetwork2020}. 
\ac{GATT} protocols consist of services which can contain multiple characteristics. In Apple's specification, two opcodes are defined to start and stop a sound if they are written to a specific characteristic. 
The characteristic is still available on sold Find My accessories, like the Chipolo ONE Spot, but the opcodes have changed. Also, the AirTag supports the same functionality, but it uses an entirely different \ac{BLE} characteristic. 

To identify the actual values of these non-owner sound characteristics, we used two methods of reverse-engineering: iOS system logs and Android Bluetooth \ac{HCI} snoop logs. 

We used an AirTag that triggered a tracking notification on iOS, then we played a sound on the device and retrieved the system logs of the \texttt{locationd} daemon at the same time. Fortunately, the logs print the characteristic that needs to be accessed and the value that needs to be written. In iOS 15.2, the AirTag only supports playing a sound but not stopping it.

On Android, we activated the Bluetooth \ac{HCI} snoop logs and played a sound on the Chipolo ONE Spot using the Tracker Detect app. The \ac{HCI} logs contain all commands that are sent to the Bluetooth chip from the host (i.e., Android) system and the responses from the chip (e.g., \ac{GATT} characteristics used). We extracted the values needed to start a sound and stop the sound. 

The discovered values to play and stop sounds are listed in \cref{tab:characteristics}. 

\section{iOS Tracking Detection}\label{apdx:ios_tracking_detection}

\begin{figure}[b]
    \centering
    \includegraphics[width=0.8\linewidth]{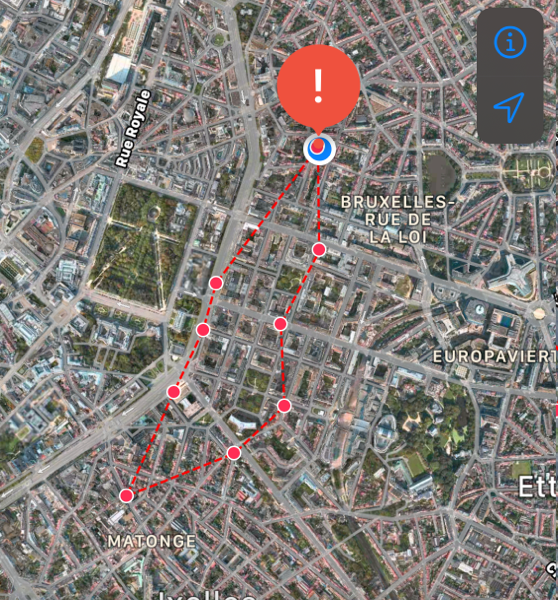}
    \caption{Screenshot of an AirTag tracking an author displayed in the Find My App.}
    \label{fig:airtag_location_path}
\end{figure}

\Cref{fig:airtag_location_path} shows a screenshot from Apple's Find My app. It shows a path constructed by several points where the iPhone has seen the AirTag as a tracking device. 

%
%
%
\begin{acronym}

\acro{BLE}{Bluetooth Low Energy}
\acro{RSSI}{received signal strength indicator}
\acro{TLV}{Type Length Values}
\acro{UWB}{Ultra-Wide Band}
\acro{HCI}{Host Controller Interface}
\acro{GATT}{Generic Attribute Profile}
\end{acronym}

\end{document}